# Direct observation of enhanced magnetism in individual size- and shape-selected *3d* transition metal nanoparticles

Armin Kleibert[*,1], Ana Balan[1], Rocio Yanes[2], Peter M. Derlet[3], C. A. F. Vaz[1], Martin Timm[1], Arantxa Fraile Rodríguez[4], Armand Béché[5], Jo Verbeeck[5], Rajen S. Dhaka[1,6,7], Milan Radovic[1,7,8], Ulrich Nowak[2], and Frithjof Nolting[1]

[1]Swiss Light Source, Paul Scherrer Institut, 5232 Villigen PSI, Switzerland

[2]Department of Physics, University of Konstanz, 78457 Konstanz, Germany

[3]Condensed Matter Theory Group, NUM, Paul Scherrer Institut, 5232 Villigen PSI, Switzerland

[4]Departament de Física de la Matèria Condensada and Institut de Nanociència i Nanotecnologia (IN2UB), Universitat de Barcelona, 08028 Barcelona, Spain

[5]EMAT, University of Antwerp, 2020 Antwerpen, Belgium

[6]Department of Physics, Indian Institute of Technology Delhi, Hauz Khas, New Delhi-110016, India

[7]Institute of Condensed Matter Physics, Ecole Polytechnique Fédérale de Lausanne (EPFL), 1015 Lausanne, Switzerland

[8]SwissFEL, Paul Scherrer Institut, 5232 Villigen PSI, Switzerland

*Corresponding author, e-mail: armin.kleibert@psi.ch

**ABSTRACT**

Magnetic nanoparticles are important building blocks for future technologies ranging from nano-medicine to spintronics. Many related applications require nanoparticles with tailored magnetic properties. However, despite significant efforts undertaken towards this goal, a broad and poorly-understood dispersion of magnetic properties is reported, even within mono-disperse samples of the canonical ferromagnetic *3d* transition metals. We address this issue by investigating the magnetism of a large number of size- and shape-selected, individual nanoparticles of Fe, Co, and Ni using a unique set of complementary characterization techniques. At room temperature only superparamagnetic behavior is observed in our experiments for all Ni nanoparticles within the investigated sizes, which range from 8 to 20 nm. However, Fe and Co nanoparticles can exist in two distinct magnetic states at any size in this range: (i) a superparamagnetic state as expected from the bulk and surface anisotropies known for the respective materials and as observed for Ni; and (ii) a state with unexpected stable magnetization at room temperature. This striking state is assigned to significant modifications of the magnetic properties arising from metastable lattice defects in the core of the nanoparticles as concluded by calculations and atomic structural characterization. Also related with the structural defects, we find that the magnetic state of Fe and Co nanoparticles can be tuned by thermal treatment enabling one to tailor their magnetic properties for applications. This work demonstrates the importance of complementary single particle investigations for a better understanding of nanoparticle magnetism and for full exploration of their potential for applications.

# I. INTRODUCTION

Magnetic nanoparticles attract a wide interest in many fields ranging from bio-medicine to energy, magnetic data storage, and spintronics [1-4]. This interest is driven by the unique magnetic phenomena which occur at the nanoscale, such as single domain states and superparamagnetism (SPM) [5]. Moreover, enhanced magnetic moments and magnetic anisotropy energies have been reported for atomic clusters and nanoparticles [6-8]. These features are of great interest for novel applications, but achieving control remains challenging and requires deeper understanding of the magnetic properties at the nanoscale. Extensive efforts have been undertaken to establish simple laws to predict size-dependent properties such as the magnetic anisotropy energy [9-12]. However, experimental validation of scalable regimes has not been achieved so far, even for the common ferromagnetic *3d* transition metals, Fe, Co, and Ni. Instead, the available literature reveals a significant scatter of magnetic properties which cannot be assigned only to particle size or environment. For instance, the magnetic anisotropy energies of Fe nanoparticles are reported to range from bulk-like to strongly enhanced values in different experiments [13-19]. Similarly, for Co nanoparticles the experimentally observed values vary over several orders of magnitude [20-25]. For Ni, the situation seems even more complex, since not only does the magnetic anisotropy energy vary, but also the magnetic moment of the particles differs in various reports [26-32]. Such variability is often assigned to shape, surface or interface effects [18,22,33]. However, an unambiguous interpretation of experimental data is difficult, since most of the reported investigations have been carried out with experimental techniques that average over large distributions of particle sizes, morphologies, and orientations. The situation might be further complicated by additional inter-particle interactions, which can largely affect ensemble properties such as magnetization curves acquired with bulk SQUID and vibrating sample magnetometry, or integrated X-ray magnetic circular dichroism (XMCD) spectroscopy [15,34,35].

In the present work we overcome these difficulties by investigating the magnetism of a large number of individual Fe, Co, and Ni nanoparticles by means of X-ray photo-emission electron microscopy (X-PEEM) together with the XMCD effect under ultrahigh vacuum conditions [15,36-39]. The magnetic properties are directly correlated with morphological information of the very same nanoparticles such as size and shape obtained by scanning electron microscopy (SEM) and atomic force microscopy (AFM). Using this unique approach, we have recently shown that *as grown* Fe nanoparticles can be found in two different states with distinct magnetic properties at any size in the range from 8 nm to 20 nm [40]. Notably, half of the particles were found in a state with strikingly high magnetic anisotropy, resulting in stable magnetism at room temperature even in the smallest investigated nanoparticles, which could be of great interest for applications where nanomagnets with high magnetic anisotropy energy and high saturation magnetisation are required. However, the high anisotropy state was found to be metastable and to relax towards a state with the (much smaller) magnetic anisotropy of bulk Fe upon thermal excitation. Further, the experiments allowed us to exclude that the unusual high magnetic anisotropy energy is due to possible surface or shape contributions to the effective magnetic energy barriers, but instead the data indicate that the enhanced energy barriers originate from metastable, structural modifications in the volume of the nanoparticles. While these data suggest that part of the controversy in the literature on the magnetic properties of Fe nanoparticles could be due to the presence of such metastable magnetic properties, important questions about the origin and nature of these observations remain open.

These questions concern particularly the presence of different crystallographic order within the investigated particle ensembles as well as thermal stability of the particle structure.

Moreover, it remained unclear whether similar magnetic behavior can be found in other *3d* transition metal nanoparticle systems as well. Finally, quantitative estimates on the impact of structural defects on the magnetic properties are needed. In this work we address these issues and demonstrate that also *as grown* Co nanoparticles exhibit a similar size-independent co-existence of nanoparticles with distinct magnetic anisotropy energies, showing that the presence of metastable states with anomalous high magnetic barrier energies is a more general phenomenon and not solely restricted to Fe. However, in contrast to Fe, the state with enhanced magnetic anisotropy in Co can be promoted by thermal annealing and thus might be of great relevance for applications. In Ni nanoparticles, uniform SPM behaviour is found at room temperature with a magnetic blocking temperature of 100 K, confirming ferromagnetic order. To address the role of the particle structure, the magnetic data are correlated with characterization obtained by means of reflection high energy electron diffraction (RHEED) and high resolution scanning transmission electron microscopy (HR-STEM). Quantitative comparison of the experimental data with theoretical model calculations, allows us to rule out that the observed variability in the magnetic anisotropy energy in Fe and Co is due to particle interactions, surface contributions or shape and size-variations. Instead, our data and quantitative estimates suggest that lattice defects within the particles are at the origin of the reported magnetic diversity and of the observed metastability. Finally, we discuss additional implications of structural defects on the magnetism of nanoparticles.

### III. EXPERIMENTAL DETAILS

The samples for the *in situ* X-PEEM experiments are prepared in three steps: (i) Au markers for particle identification in complementary microscopy investigations are lithographically prepared on Si(100) wafer substrates passivated with a native $SiO_x$ layer, see Fig. S-1 of the Supplementary Material (SM) [41,42]. (ii) Upon introduction into the ultrahigh vacuum (UHV) surface preparation system (SPS) (base pressure $\leq 5\times10^{-10}$ mbar), the substrates are treated to remove adsorbates such as water which originate from exposure to the ambient atmosphere. In the case of the Fe nanoparticles the substrates were cleaned by mild sputtering with argon ions (kinetic energy: 500 eV, argon pressure: $5\times10^{-5}$ mbar, duration: 20 min), while for the Co and Ni nanoparticles, the substrates were thermally annealed *in situ* for 30 mins at about 525 K in the SPS. The SPS is directly attached to the PEEM instrument. For the RHEED studies, plain Si(001) wafers with the native $SiO_x$ surface layer are used. The wafers are annealed in the UHV RHEED system (base pressure: $\leq 5\times10^{-9}$ mbar) at a temperature of about 525 K until the pressure in the chamber recovers (after an initial increase) and the recorded RHEED pattern indicate a clean and flat $SiO_x$ surface. (iii) Finally, the nanoparticles are deposited onto the prepared substrates using an arc cluster ion source (ACIS), which is attached to the SPS [43-45]. For RHEED and X-PEEM investigations all samples are transferred under UHV conditions. This approach allows us to study the pristine magnetic properties of the nanoparticles.

In the ACIS, the nanoparticles are formed by condensation of metal vapor in a carrier gas consisting of a He/Ar mixture [43]. The metal vapor is generated by means of arc erosion from respective metal targets with a purity of 99.8%. An electrostatic quadrupole deflector is used to deflect a beam of mass-filtered nanoparticles onto the previously prepared Si substrates which are held either directly in the SPS or in a vacuum suitcase (base pressure $\leq$ $5\times10^{-9}$ mbar) attached to the SPS. A gold mesh placed in the nanoparticle beam path is used to measure the flux of the electrically charged particles during deposition and to control the final particle density on the substrates. For the X-PEEM investigations we choose a low particle density (a few nanoparticles per $\mu m^2$) to avoid magnetic dipolar interactions between the nanoparticles and to enable single particle resolution in the X-PEEM experiments (the particle-particle distance should be larger than 200 nm) [38]. For the RHEED experiments we choose a higher particle density of about 30 nanoparticles per µm$^2$ in order to obtain a sufficient signal-to-noise ratio in the diffraction data. At this coverage, agglomeration of the particles on the substrate is still avoided as confirmed by subsequent SEM images, so that also the RHEED data reflects the properties of an ensemble of isolated nanoparticles. Finally, samples with a particle density of a few tens of particles per µm$^2$ for *ex situ* HR-STEM investigations were deposited in the SPS. Commercially available 10 nm SiN membranes were used as substrates as-received (TEMwindows.com). During nanoparticle deposition the pressure temporarily increases to about 5×10$^{-6}$ mbar due to the presence of the Ar/He carrier gas, but recovers to the respective base pressure within a few minutes after deposition. For the present work the cluster source operation parameters as well as the mass-filter settings are held constant for all samples. This ensures similar growth, selection, and landing conditions in all experiments, with the kinetic energy of the particles prior to the impact on the substrate smaller than 0.1 eV/atom [44]. With these settings, the deposition takes place under so-called soft landing conditions, where no fragmentation of the particles or damage to the substrate is expected [46,47].

The crystallographic structure, the orientation of the deposited nanoparticles with respect to the substrate, as well as the thermal stability of the particles and the substrate, are determined by RHEED measurements [48,49]. The RHEED experiments are carried out with electrons with a kinetic energy of 35 keV at grazing incidence. This geometry enables one to investigate the quality of the substrates and the deposited nanoparticles simultaneously [48,50]. Data is recorded using a charge coupled device camera attached to the phosphor screen of the instrument. The temperature is set by means of resistive heating of a Si wafer piece under the sample. The sample temperature is read by a pyrometer (Maurer GmbH, Typ: KTR 1075-1).

The *in situ* magnetic characterization of the samples is carried out using the PEEM (Elmitec GmbH) at the Surface/Interface: Microscopy (SIM) beamline of the Swiss Light Source [51]. The base pressure in the PEEM chamber is $< 5\times10^{-9}$ mbar for the Fe nanoparticle experiments and $< 5\times10^{-10}$ mbar for the Co and Ni nanoparticle investigations. For X-PEEM imaging the samples are illuminated with polarized mono-chromatic synchrotron radiation. The nanoparticles are visualized by means of elemental contrast maps, which are obtained by recording two images at a given sample site: first, a so-called "edge"-image is recorded with the photon energy resonantly tuned to the respective element-specific $L_3$ X-ray absorption edge. Then, a second so-called "pre-edge"-image is recorded with the photon energy tuned a few eV below the $L_3$ X-ray absorption edge energy. Pixel-wise division of the "edge"- and "pre-edge"-images finally yields the elemental contrast map, which reveals the nanoparticles as bright spots on the image, cf. Figs. 1(a) – 1(c) [41]. The photon energies used in the resonant excitation of the $L_3$ X-ray photo-absorption edges for the "edge"-images are 708 eV for Fe, 778 eV for Co, and 852 eV for Ni. The photon energies used for recording the "pre-edge"-images are 703 eV for Fe, 773 eV for Co, and 847 eV for Ni. A typical measurement sequence consists of averaging 10 individual frames with 1 s integration time each per photon energy from which a sequence of 10 elemental contrast maps is obtained. This sequence is then corrected for possible sample drift and finally averaged to yield elemental contrast images such as shown in Figs. 1(a) – 1(c).

The magnetic properties of the particles are probed using the XMCD effect [52]. The latter gives rise to a magnetization- and helicity-dependent X-ray absorption cross section when tuning the photon energy resonantly to the $L_3$ absorption edge of the nanoparticles [52]. Magnetic contrast maps are obtained by pixelwise division of two X-PEEM images recorded with circularly polarized light of opposite helicity, $C^{\pm}$. In these maps, particles will exhibit a gray tone contrast ranging from black to white, depending on the projection of their magnetic moments onto the propagation vector of the X-ray beam, see Fig. S-2 of the SM [42]. The time resolution of the present X-PEEM investigations is limited to $\tau_x = 20$ s, corresponding to a single pair of averaged $C^+$ and $C^-$ images with a total acquisition time of 10 s each. For the data in Figs. 1(d) – 1(f) a sequence of 20 magnetic contrast maps was acquired, drift-corrected and averaged.

The spatial resolution of the X-PEEM experiments is limited to 50 - 100 nm, thus largely exceeding the size of the particles. The nanoparticle size and morphology is therefore investigated by means of complementary SEM (Zeiss Supra VP55) and AFM (Veeco di 3100) after the *in situ* magnetic characterization. During transfer to the SEM the samples are exposed to ambient air. The lithographic marker structures allow one to identify the very same particles in X-PEEM, SEM and AFM, see Fig. S-1 of the SM [42]. The SEM is used to investigate the lateral morphology of the individual particles and to exclude close-lying particles, agglomerates and irregular, dendritic structures from the analysis. Since the spatial resolution is limited to 2 – 3 nm, an accurate size determination with SEM is not possible. Therefore, AFM is used to determine the height of the particles which serves as a measure of their size [53]. Finally, the morphology of the particles (exposed to air) was investigated by means of a HR-STEM (FEI Titan$^3$ equipped with Cs probe corrector) with high-angle annular dark-field (HAADF) imaging.

**III. RESULTS**

**A. *In situ* magnetic characterization with single particle sensitivity**

X-PEEM elemental and magnetic contrast maps of *as grown* Fe, Co, and Ni nanoparticle samples recorded at room temperature (RT) are shown in Fig. 1. Bright spots in the elemental contrast maps in Figs. 1(a) for Fe, 1(b) for Co, and 1(c) for Ni indicate individual nanoparticles within the extended samples. As a guide to the eye, a number of particles are highlighted in Fig. 1 with white solid and dashed circles, respectively. Magnetic contrast maps of the same sample areas in Figs. 1(d) and 1(e) reveal that about half of the Fe and Co nanoparticles exhibit magnetic contrast ranging from black to white, for instance the particles highlighted with the solid white circles. The actual magnetic contrast of an individual nanoparticle depends on the orientation of its magnetic moment relative to the propagation vector of the impinging X-rays (see II.C). The presence of magnetic contrast further indicates a magnetically blocked state, i.e. a stable magnetic moment orientation over time periods longer than the measurement acquisition time of ~20 s in the present experiments [40]. Such particles are referred to as FM in this work. A detailed analysis of the magnetic contrast intensity distribution of a large number of FM nanoparticles reveals a random orientation of their magnetic moments reflecting the stochastic nature of the nanoparticle deposition process [for details of the analysis cf. the SM [42], in particular Figs. S-3 and S-4]. The lack of magnetic contrast in the other half of the nanoparticles can be assigned to thermally excited magnetic moment reversals in these nanoparticles at a rate higher than the measurement acquisition time, i.e. to superparamagnetic behavior, see also Ref. [40]. Such particles are referred to as SPM in this work. A number of SPM particles are highlighted with dashed white circles in Fig. 1. The absence of magnetic contrast in the Ni nanoparticle sample, Fig. 1(f), shows that all Ni nanoparticles are SPM at RT, with magnetic contrast occurring at about 100 K (not shown).

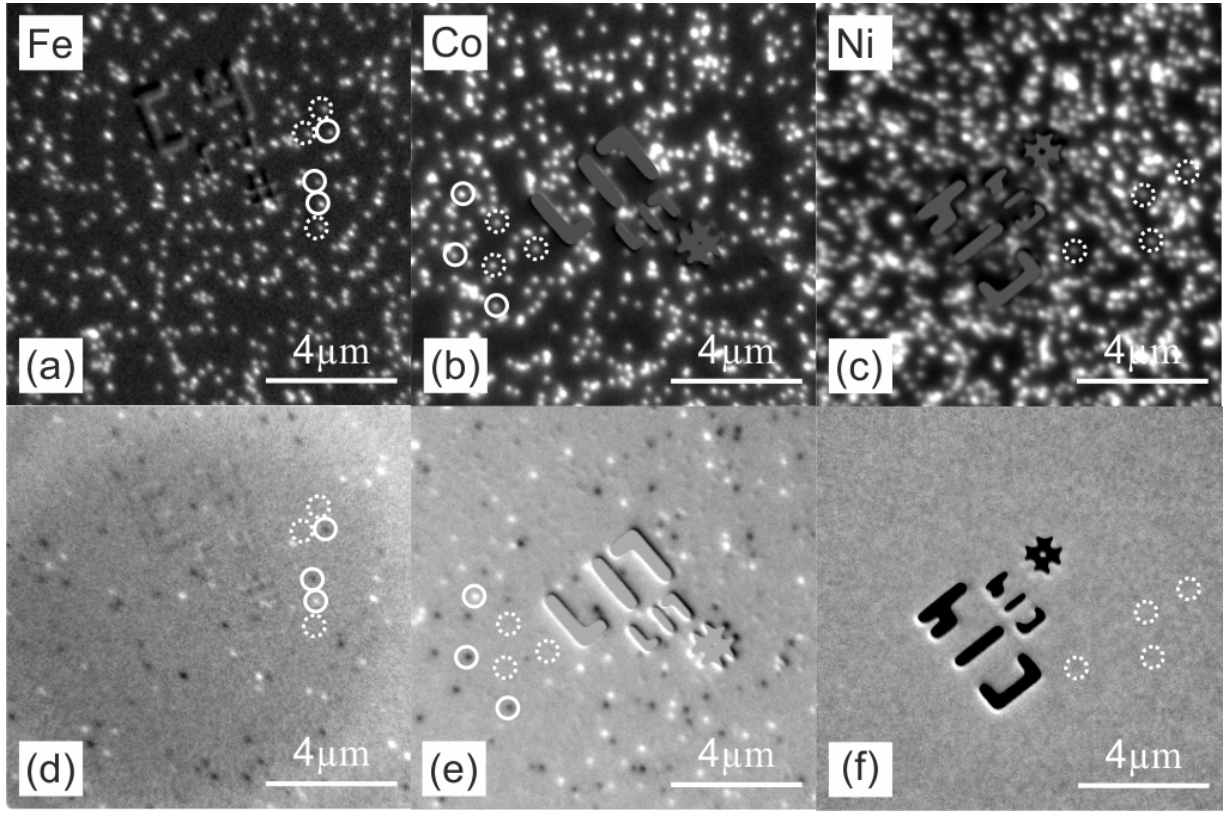

FIG. 1. (a)-(c) Elemental contrast maps of *as grown* (a) Fe, (b) Co, and (c) Ni nanoparticles at RT. (d)-(f) Respective magnetic contrast maps. Examples of particles in superparamagnetic and magnetically blocked states are highlighted with dashed and solid circles, respectively.

**B. Structural characterization by means of RHEED**

The crystallographic structure impacts on both the magneto-crystalline anisotropy and the shape of the particles and needs to be addressed experimentally, in particular, since the atomic lattice structure of nanoparticles can vary depending on the preparation technique and growth conditions [54]. For instance, in the literature Co and Ni nanoparticles were reported to exist in various structures ranging from hexagonal closed packed (hcp), primitive cubic, to face centered cubic (fcc) [27,55,56], while Fe has been stabilized in body centered cubic (bcc) and fcc at the nanoscale [48,57,58]. RHEED data taken for *as grown* Fe, Co, and Ni nanoparticles are shown in Figs. 2(a), 2(b), and 2(c) and exhibit characteristic Laue ring patterns from which the lattice structure can be deduced. For the Fe nanoparticles the Laue pattern is consistent with that of the bcc lattice [48]. The presence of diffraction rings confirms a nearly random crystallographic orientation of the particles upon deposition, which agrees with the random orientation of the magnetic moments of the FM particles, cf. Fig. 1(d). A texture is found on top of the (200) and the (110) rings. This indicates that the Fe nanoparticles preferentially rest with (100) and (110) facets parallel to the substrate surface [48]. This observation agrees with the expected shape according to a Wulff construction of mono-crystalline bcc Fe nanoparticles given by a truncated dodecahedron exhibiting 6 (100) and 12 (110) facets [59,60], see also the inset in Fig. 6(a). The RHEED data of the Co and Ni nanoparticles show that they crystallize in the fcc structure, in accordance with other reports on gas phase grown systems in the present size range [56,61]. Again, we find a texture in the two lower index rings, (111) and (200), which suggests a preferred resting on (111) and (100) surface facets being consistent with the Wulff shape of mono-crystalline fcc nanoparticles given by truncated cuboctahedra with 8 (111) and 6 (100) surface facets [59] and being schematically depicted in the insets of Figs. 6(b) and 6(c). RHEED data taken at larger scattering angles further reveal an intact and flat, amorphous $SiO_x$ surface layer after the deposition of the nanoparticles (not shown).

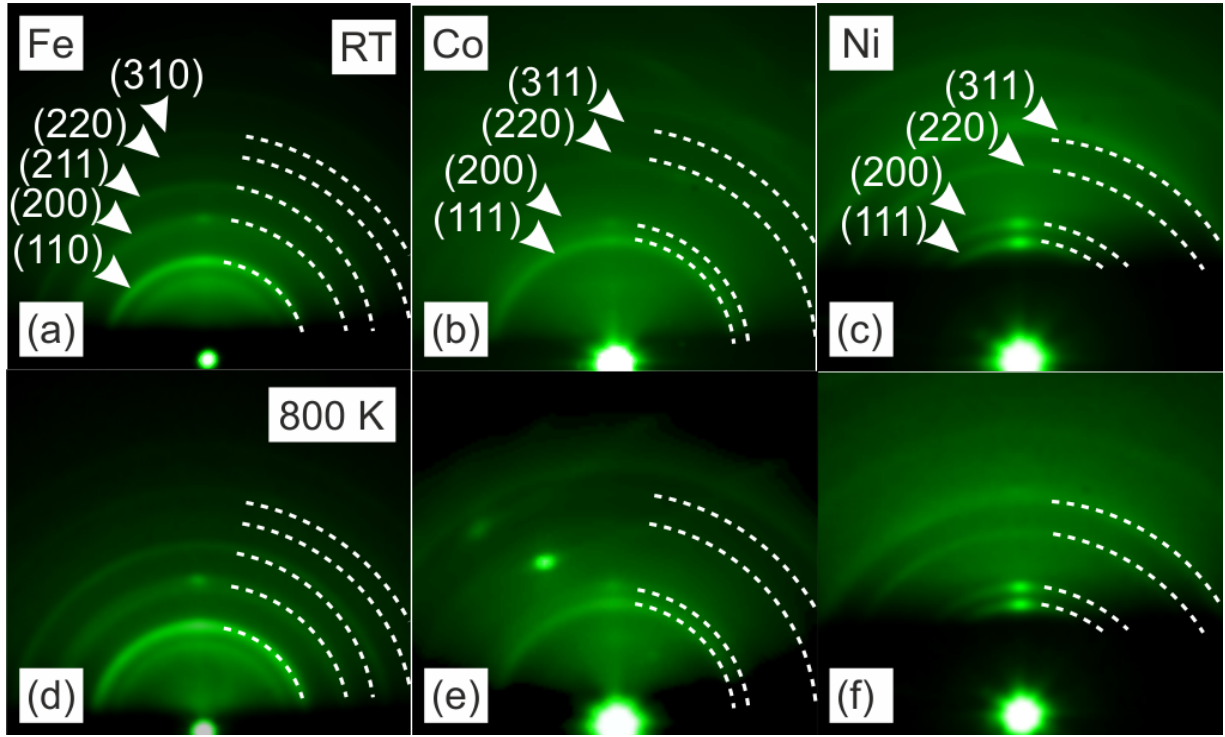

FIG. 2. (Color online) RHEED patterns for *as grown* (a) Fe, (b) Co, and (c) Ni nanoparticles obtained at RT. (d) – (f) corresponding diffraction patterns recorded at 800 K. Dashed arcs are guides to the eye and highlight the detected diffraction rings. The corresponding Miller indices are indicated by the small arrows [48,49,62].

When increasing the sample temperature to 800 K, no change is observed for all samples, cf. Figs. 2(d) - 2(f), suggesting a high structural and chemical stability of the nanoparticles and of their interface with the substrate. Only when approaching the thermal decomposition temperature of the $SiO_x$ surface layer of about 1050 K, do the diffraction rings disappear and discrete diffraction spots occur (not shown) [63]. This observation indicates a chemical reaction of the particles with the exposed Si(001) substrate at higher temperatures. In turn, the absence of such diffraction spots in the RHEED pattern in the *as grown* samples, further confirms the non-destructive nature of the present soft-landing nanoparticle deposition, which not only avoids fragmentation of the nanoparticles, but also preserves the ultrathin $SiO_x$ surface layer (~1.5 nm thick) [46,47].

**C. *Ex situ* morphology characterization of the nanoparticles**

*Ex situ* scanning electron microscopy (SEM) images of the samples are presented in Figs. 3(a) – 3(c). All samples reveal a variety of shapes which range from highly symmetrical and compact particles to dendritic structures or agglomerates of particles. This variety is assigned to growth kinetics in the cluster source used here for the deposition of gas phase grown nanoparticles. The growth depends on a number of parameters, such as temperature, nucleation density, and cooling rates [47,64]. With the chosen cluster source and deposition settings, which are identical for all reported experiments here, we observe a relatively high fraction (about 2/3) of compact, i.e., nearly spherical or cubic, nanoparticles of Fe and Ni [cf. the insets in Figs. 3(a) – 3(c)]. In contrast, the Co samples show a significantly lower fraction of regular particles (about 1/3). Using substrates with gold marker structures allows us to perform detailed SEM investigations on the very same nanoparticles, which were previously magnetically characterized by means of X-PEEM, see SM [42] in particular Fig. S-1. Specifically, the SEM data are used to select only nearly spherical or cubic nanoparticles for the analysis of their magnetic properties for all three investigated systems Fe, Co, and Ni. Typical examples are shown in the insets to Figs. 3(a) – 3(c). Particles with more complex morphology are not further considered in the present work. This choice is made to facilitate the comparison with the model calculations discussed below.

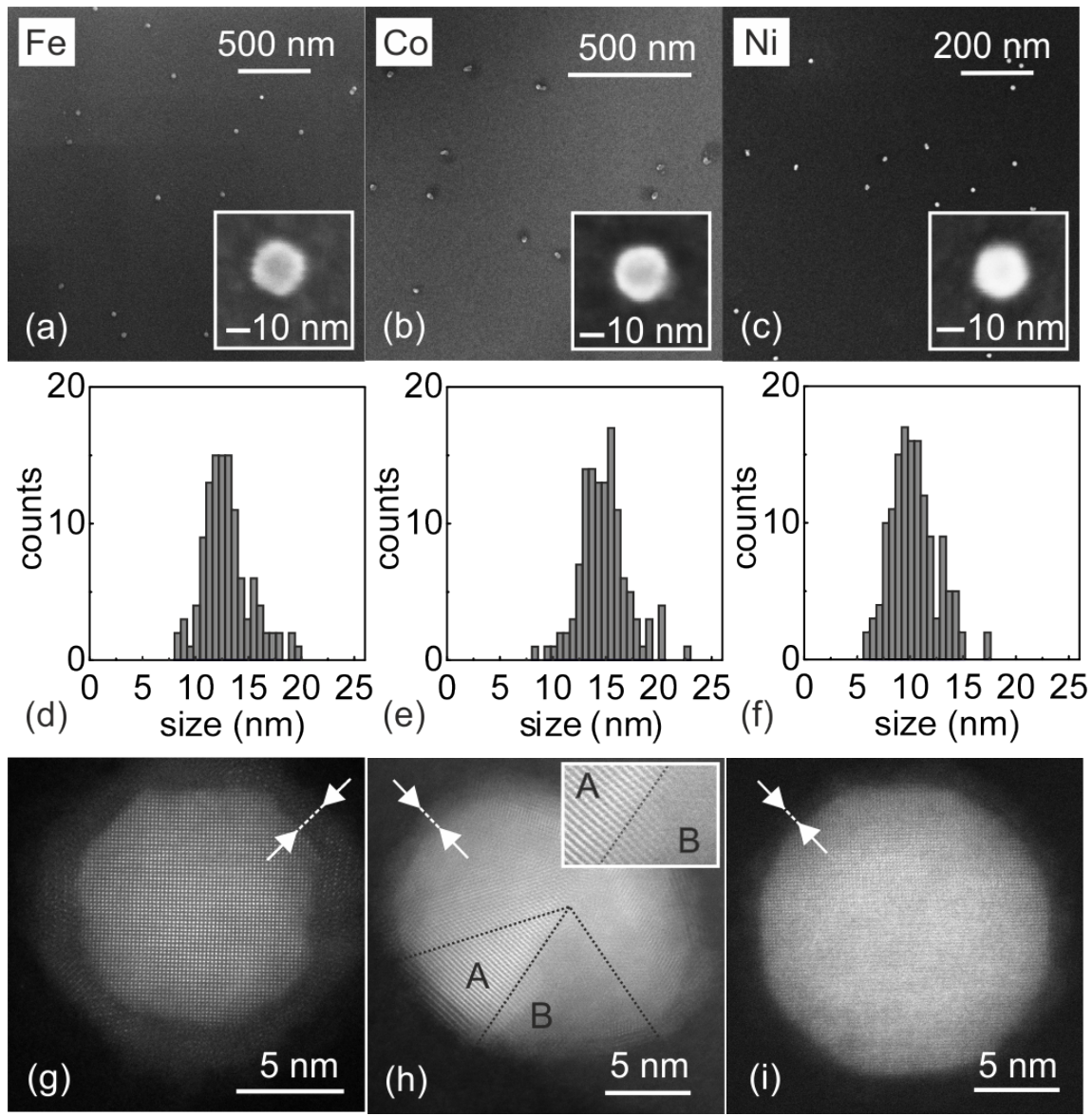

FIG. 3. (a)-(c) SEM micrographs of the (a) Fe, (b) Co, and (c) Ni nanoparticle samples recorded after the PEEM experiments. The insets show magnified images of representative particles with a regular shape selected for further analysis. (d)-(f) AFM height distribution obtained from the Fe (d), Co (e), and Ni (c) particles selected based on the SEM data. (g)-(i) HAADF HR-STEM images of Fe (g), Co (h), and Ni (i) nanoparticles. The arrows indicate the thickness of the oxide shell which has formed due to ambient air exposure. The black dotted lines in panel (h) indicate two zones "A" and "B", respectively, with different crystallographic orientations.

The size of the selected nanoparticles is obtained by measuring their height using atomic force microscopy (AFM) [65]. Figs. 3(d) – 3(f) show the actual size distributions for the considered Fe, Co, and Ni nanoparticles as obtained from AFM analysis. The sizes are nearly the same for all three samples and range from about 8 to about 20 nm as expected from the identical deposition conditions. The mean values are about 12 nm. However, all particle heights are affected by the formation of an oxide shell, which occurs upon ambient air exposure after the X-PEEM investigations and during transfer to the SEM and AFM instruments. The oxide shell thickness is characterized by means of HR-STEM investigations carried out on air exposed reference samples. The HR-STEM images confirm the presence of an oxide shell formed in all samples, cf. Figs. 3(g) – 3(i). The data suggests that Ni nanoparticles possess the thinnest oxide shell with a thickness of about 1 nm, while Co and Fe nanoparticles develop an oxide shell of 2 to 3 nm in agreement with previous studies [60,66,67]. The actual particle sizes in the ultrahigh vacuum during the X-PEEM and RHEED investigations might therefore be smaller by about 1 - 2 nm when compared to the *ex situ* AFM results presented in Figs. 3(d) – 3(f). In what follows we refer always to the AFM data without correction of the oxide shell thickness. The HR-STEM data further confirm that the Co particles exhibit a larger variety of shapes as suggested by SEM and further reveal a number of twinned or polycrystalline particles, as can be seen for instance in Fig. 3(h).

**D. Correlation of magnetic properties and particle size**

Finally, our unique complementary microscopy approach is used to correlate magnetism and size of a large number of individual, shape-selected nanoparticles. This is achieved by analyzing the size distribution of FM and SPM nanoparticles, separately. The results of this analysis are shown in Figs. 4(a) – 4(c) for all three systems. Surprisingly, the data reveal the same signature for Fe and Co nanoparticles, i.e., a nearly equal coexistence of SPM and FM nanoparticles, irrespective of size. In contrast, Ni nanoparticles are only found in the SPM state at RT. [Note that summing up the FM and SPM nanoparticle contributions in Fig. 4 yields directly the magnetically unresolved size distributions obtained by AFM as shown in Figs. 3(d) – 3(f).]

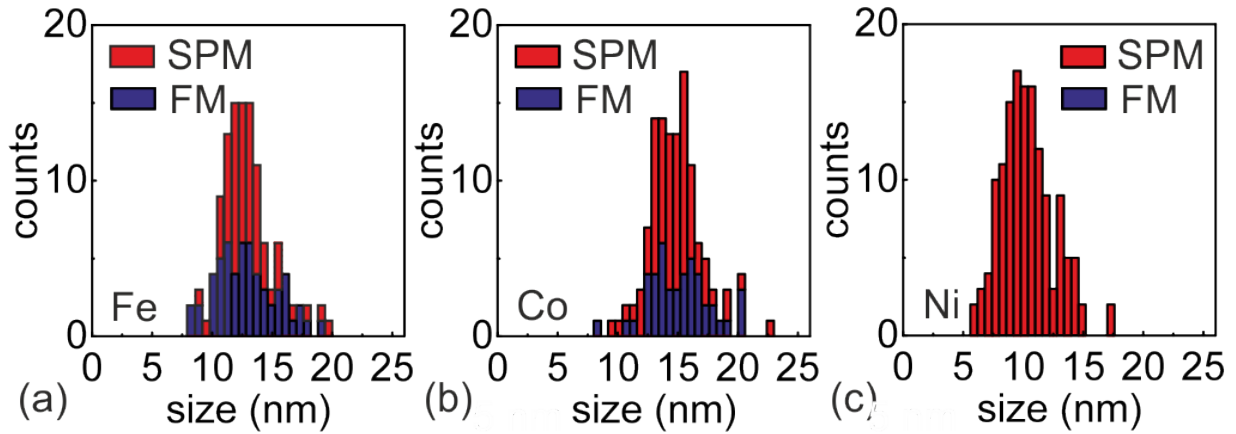

FIG. 4. (Color online) Relative fraction of SPM (red) and FM (blue) nanoparticles as a function of the particle size for *as grown* (a) Fe, (b) Co, and (c) Ni nanoparticles at RT.

**E. Thermal stability of the magnetic properties**

Due to their different lattices, bcc Fe and fcc Co nanoparticles are expected to show significantly distinct magnetic properties. In particular, the magneto-crystalline anisotropy energy of bulk fcc Co is known to result in magnetic energy barriers that are about two times smaller than that of bcc Fe. Therefore, it is a remarkable observation that both systems exhibit the same size-independent coexistence of FM and SPM nanoparticles at RT. To further understand this behaviour we studied the effect of thermal annealing on the magnetic properties of the Co nanoparticles as well as on those of the Ni nanoparticles. For the case of Fe we had recently demonstrated by means of *in situ* X-PEEM investigations that the FM state is metastable and can relax towards the SPM state [40]. In particular, it was found that all Fe nanoparticles lose their magnetic contrast when raising the sample temperature to 420 K. When cooling the sample to RT, the initial magnetic contrast is not recovered, indicating that all initially FM Fe nanoparticles undergo an irreversible transition to the SPM state [40].

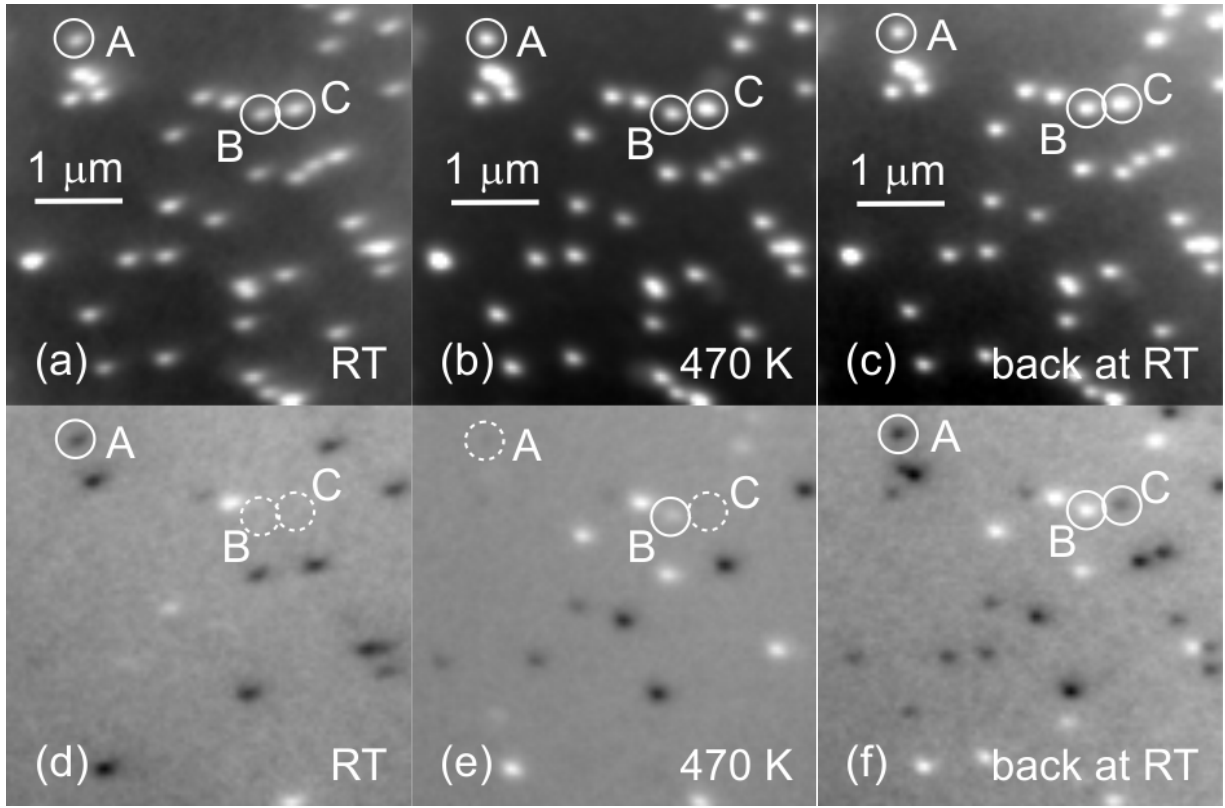

FIG. 5. Elemental and magnetic contrast maps of *as grown* Co nanoparticles (a) and (d), at RT, (b) and (e) at 470 K, (c) and (f) after cooling back to RT. Three particles labelled A, B, and C are highlighted for discussion.

The results of a similar study for Co are shown in Fig. 5. We find that only a relatively small number of the initially FM particles lose their magnetic contrast when rising the temperature to 470 K, for example, particle A in Figs. 5(d) and 5(e). Instead most initially FM particles remain in a magnetically blocked state. Moreover, some particles, initially without magnetic contrast at RT, surprisingly display magnetic contrast at 470 K, e.g., nanoparticle B in Figs. 5(d) – 5(f). These particles remain in the FM state when cooling back to RT. In addition, a number of nanoparticles become FM at RT only after the thermal cycle, such as nanoparticle C in Figs. 5(d) – 5(f). This behaviour suggests thermally induced irreversible transitions from SPM to FM states. Finally, some particles of type A do not change their properties and exhibit a reversible transition from FM behaviour at RT to SPM behaviour at 470 K. Hence, these observations reveal a remarkable difference between the Fe and the Co nanoparticles. A similar procedure has no effect for the magnetic state of the Ni nanoparticles, i.e., the entire ensemble remains SPM before, during, and after thermal annealing (not shown).

## IV. DISCUSSION

### A. Atomic level simulation of magneto-crystalline and effective surface anisotropy contributions for spherical nanoparticles

In order to evaluate the experimental findings we have performed advanced atomistic model calculations of the magnetic energy barriers $E_m$ of defect-free bcc Fe, fcc Co and fcc Ni nanoparticles, which also include the effect of non-collinear surface spin configurations due to the Néel-type surface anisotropy $K_s$ [68,69]. The simulations consider classical spins distributed over the lattice sites of spherical model particles with diameter $D$. The magnetic properties of the particles are described by an anisotropic Heisenberg Hamiltonian [69]. The effective energy landscapes of the many-spin particles are then evaluated using the Lagrangian multiplier method as described in Refs. [68,70,71]. The exchange constants are chosen to reproduce the bulk Curie temperatures using the classical spectral density method [72]. The magneto-crystalline anisotropy energy density is given by $e_{MCA} = K_1(\alpha_1^2\alpha_2^2+\alpha_2^2\alpha_3^2+\alpha_3^2\alpha_1^2) + K_2\alpha_1^2\alpha_2^2\alpha_3^2$, where $\alpha_i$ are the direction cosines and $K_{1,2}$ the tabulated first and second order anisotropy constants of the respective bulk material at room temperature. For ease of discussion we use the constants on a per atom basis, which are for bcc Fe: $K_1$ = 3.8 μeV/atom, $K_2$ = 0.008 μeV/atom, for fcc Co: $K_1$ = -3.8 μeV/atom, $K_2$ = -0.77 μeV/atom, and for fcc Ni: $K_1$ = -0.28 μeV/atom and $K_2$ = -0.12 μeV/atom. All other parameters are listed in the table given in the SM [42].

Before we discuss the effect of the surface anisotropy, we consider first the magnetic energy barriers due to the magneto-crystalline anisotropy. For the discussion we refer to the schematic Wulff-shaped particles as shown in the insets of Figs. 6(a) – 6(c), which provide a better visualization of the crystallographic directions when compared to spheres. With the given values of $K_{1,2}$ we find that for the bcc Fe (fcc Co and Ni) particles the easy axes are along <100> (<111>) and the hard axes are along <111> (<100>). In Figs. 6(a) – 6(c) the easy (hard) axes are indicated by black (red) arrows next to the schematics. The lowest energy path for the magnetic moments to switch from one easy axis to another is along the semi-hard axis, which is along the <110> direction for all systems. The corresponding path for the magnetization is schematically indicated by the dotted arcs in the insets of Figs. 6(a) – 6(c). The respective magnetic energy barriers amount to 0.8 μeV/atom for bcc Fe, to 0.3 μeV/atom for fcc Co, and to 0.03 μeV/atom for fcc Ni. Finally, the total magnetic energy barriers $E_m$ as a function of $D$ are shown by the black lines in Figs. 6(a) – 6(c). For comparison with the present experiments, the dashed horizontal lines in all panels of Fig. 6 indicate the energy barriers above which a FM state is observed in the present experiments (for the calculation see the SM [42] and Ref. [73]). Values of $E_m$ below that threshold result in SPM behavior. The data in Figs. 6(a) – 6(c) clearly show that for all investigated materials and sizes the magneto-crystalline anisotropy alone would result in SPM states.

The effect of the additional surface anisotropy due to spin non-collinearities at the surface is shown in Figs. 6(d) – 6(f) as a function of the ratio $K_s/K_1$. We note that the high symmetry of spherical (as well as of Wulff-shaped) nanoparticles leads to a cancellation of the first order Néel surface anisotropy, and thus no surface effect is expected if non-collinear surface spin configurations are neglected [68]. The present calculations allow us to address this issue and to evaluate the actual contribution of $K_s$ to the magnetic energy barriers of spherical nanoparticles. Since $K_s$ is not a priori known and can significantly differ between various experimental reports, we have chosen to consider the range $0 < |K_s/K_1| < 800$ in order to cover a large range of experimentally determined surface anisotropies deduced from thin film studies [74-79]. Note that for calculating $K_s/K_1$ also $K_s$ is considered on a per atom basis. For comparison, ensemble measurements on nanoparticles have suggested $|K_s/K_1| \sim 300$ for Fe and $|K_s/K_1| \sim 600$ for Co [14,22]. Calculations are carried out for two particle sizes, 8 (red symbols) and 12 nm (black symbols). The data reveal that a sizeable enhancement of $E_m$ is possible for Fe and Co nanoparticles when $|K_s/K_1| > 500$, while for Ni no enhancement is found for $|K_s/K_1| < 800$. Despite the enhancement for Fe and Co, FM states are not induced by the considered surface contributions. We may note that for Co and Ni the surface anisotropy actually counteracts the magneto-crystalline anisotropy and thus initially reduces the magnetic energy barrier before it becomes the dominant contribution for higher values of $K_s$. For Co we find further that the magnetic energy landscape changes significantly for $K_s > 250 K_1$ due to the increasing surface anisotropy, cf. grey shades in Fig. 6(e). As a result the magnetic easy axes reorient and for sufficient large values of $K_s$, the easy axes will change to <100>.

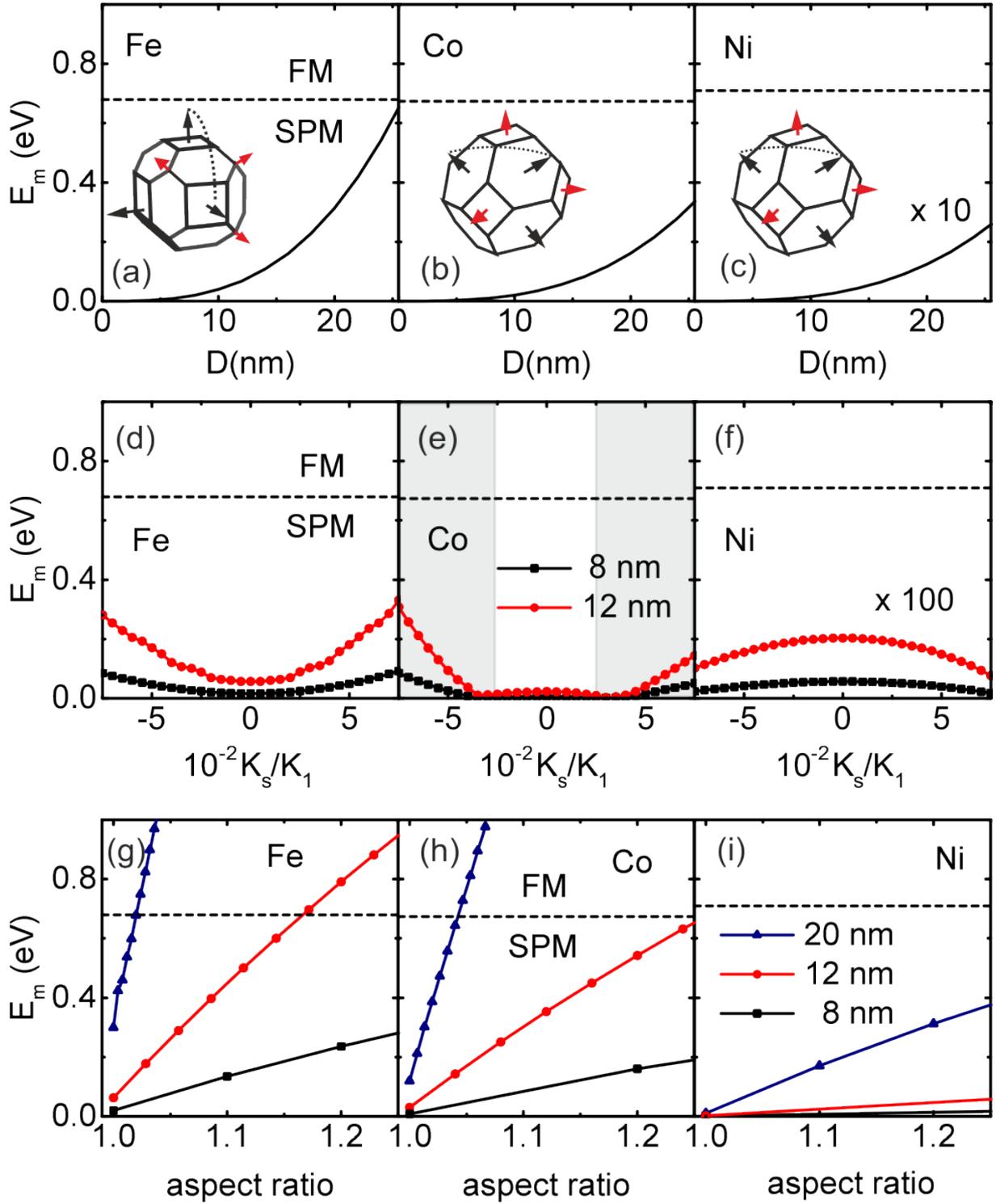

FIG. 6. (Color online) (a)-(c): Size-dependent magnetic energy barriers as given by the magneto-crystalline anisotropy energy of bulk Fe (bcc), Co (fcc), and Ni (fcc). The insets display nanoparticle shapes as predicted by the Wulff theorem for mono-crystalline bcc Fe and fcc Co and Ni, respectively. The red (black) arrows indicate the magnetic hard (easy) axes

of the particles as given by the magneto-crystalline anisotropy energy. The dotted arcs in the insets indicate the low energy barrier pathway to flip the magnetization coherently from one easy axis to another. (d)-(f) Magnetic energy barriers including surface anisotropy contributions as discussed in the text. The grey shaded areas in case of Co indicate regions where a spin reorientation occurs due to dominant surface anisotropy. (g)-(i) Magnetic energy barriers due to magneto-crystalline contributions and shape anisotropy as a function of aspect ratio. The dashed lines indicate the magnetic energy barriers above which the relaxation time is larger than 20 s. Particles with equal or higher $E_m$ would appear as FM in our experiments.

**B. Shape anisotropy contributions**

The calculations show that the experimentally observed FM states in Fe and Co nanoparticles are not due to magneto-crystalline and surface anisotropy contributions. However, another contribution to the magnetic energy barrier can arise from deviations of the particle shape from the ideal spherical or highly symmetrical Wulff construction. The resulting dipolar stray field energy can cause a sizeable magnetic shape anisotropy, which can dominate over the other contributions [16]. Figs. 6(g) – 6(i) show the calculated magnetic energy barriers for three selected particle sizes (8, 12, and 20 nm) as a function of the aspect ratio of prolate Fe, Co, and Ni ellipsoids, respectively. The calculations were performed as described in Ref. [80] using bulk values for the saturation magnetization $M_s$. The particle volume is kept constant for the different aspect ratios. In Figs. 6(g) - 6(i) the shape-induced energy barrier of the nanoparticles has been added to the respective magneto-crystalline contributions shown in Figs. 6(a) – 6(c). The shape anisotropy scales with the square of the magnetization and the corresponding energy barrier strongly increases with particle size as well as with the value of $M_s$ of the respective material. In case of bcc Fe, which has the highest $M_s$ (1702.6 emu/cm$^3$) among the materials studied here, even smaller deviations from spherical geometry result in a significant enhancement of the magnetic energy barrier [81]. The situation is similar for Co, but the energy barriers are somewhat reduced when compared to Fe due to the smaller $M_s$ of fcc Co (1428.6 emu/cm$^3$) [81]. The saturation magnetization of fcc Ni is 510.3 emu/cm$^3$ being almost less than a third of that of Fe and Co [81]. Accordingly, for the Ni nanoparticles the shape-related contributions result in much smaller magnetic energy barriers. For comparison with the experimental data the SEM investigations described above give an upper limit for the aspect ratio of 1.15 for the selected particles. According to Figs. 6(g) and 6(h), shape anisotropy-induced FM states might be therefore possible for Fe and Co nanoparticles with sizes of 12 nm and above. For Ni the shape anisotropy results in SPM behaviour at all sizes, as experimentally observed. These conclusions remain the same even when adding the surface contributions shown in Figs. 6(d) – 6(f).

## C. Role of structural defects

The calculations show that surface and shape anisotropy contributions can indeed result in a sizeable enhancement of the magnetic energy barriers of nanoparticles. It also follows that the sensitivity of the shape anisotropy to relatively small variations of the particle morphology can lead to a sizeable diversity of magnetic energy barriers even in mono-disperse nanoparticle samples. Taking into account the large shape distribution observed in many experiments, these effects will certainly contribute to the dispersion of the magnetic anisotropy energies reported in the literature. However, the calculations provide no explanation for the presently observed FM states in Co and Fe nanoparticles with size below 12 nm. In addition, for the Fe nanoparticles, recent experiments provide further evidence that the FM states even for the larger nanoparticles cannot be explained by a combination of shape, surface and magneto-crystalline anisotropy contributions [39,82]. Similarly, the metastability of the magnetic energy barriers of the Fe and Co nanoparticles hints at an additional, sizeable and variable contribution to the total magnetic energy barriers. In what follows, we argue that the origin of such phenomena lies in the presence of lattice defects such as dislocations or stacking faults. Such defects may arise from particle growth kinetics which can result in complex structures [83]. Experimental evidence for such defects is provided by the HR-STEM investigations of the Co nanoparticles in the present work as well as reported in the literature [56,84], suggesting that such structural defects are abundant in *3d* transition metal nanoparticles. Lattice defects may thus contribute to the magnetic properties in many experiments, although their effects are rarely discussed [32,84,85]. In the following, we will consider the most common lattice defects and demonstrate that such defects explain not only the metastable magnetism and the exceptionally high magnetic anisotropy energies found in Fe and Co nanoparticles, but can further result in unexpected magnetic order and spin structures in nanoparticles. Finally, some of these defects may even give rise to novel phenomena such as magnetic chirality effects in magnetic nanoparticles, which might be of interest for future applications.

Structural characterization of the *as grown* Fe nanoparticles by means of RHEED [see Fig. 2(a)] indicates solely bcc lattice symmetry, while X-PEEM reveals an almost equal amount of SPM and FM nanoparticles [see Figs. 1(d) and 4(a)]. Thus, it can be ruled out that the FM states are related to other crystal structures such as fcc-type Fe nanoparticles [86]. Rather, the data yield that FM and SPM nanoparticles are structurally very similar. This is further supported by the observation that the RHEED pattern exhibit also no noticeable changes upon thermal annealing up to 800 K, while the FM nanoparticles clearly undergo an irreversible transition to SPM behaviour upon annealing to 470 K [40]. These findings show that the transition from FM to SPM is not related with a structural phase transition, but that the distinct magnetic properties of SPM and FM nanoparticles are associated with smaller, metastable modifications of the bcc crystal lattice such as structural defects, which are typically dislocations, twinning and stacking faults in metallic nanoparticles [83,87-89]. As we show below each type of these defects is expected to have a significant and specific impact on the magnetic properties of nanoparticles.

Dislocations are common defects in bulk bcc Fe and known to cause sizeable inhomogeneous strain fields around the dislocation core. The strain fields give rise to local magneto-elastic anisotropy energy contributions and are a source of pinning sites to magnetisation reversal [90]. Continuum mechanical calculations for bcc Fe predict for instance for a single edge (screw) dislocation within a {112} slip plane a uniaxial magnetic anisotropy with energy barriers of about 43 (10) µeV/atom at a distance of 0.5 nm from the dislocation core [91]. These values are clearly much larger than the magnetic energy barrier of 0.8 µeV/atom due to the magneto-crystalline anisotropy energy of the Fe bcc lattice, and thus a dislocation locally increases the magnetic energy barriers in Fe. In the bulk, strain fields associated with dislocations extend usually over a distance of a few 100 nm. Thus, when present in a nanoparticle with a size between 8 and 20 nm, the strain field and the associated magneto-elastic anisotropy of a dislocation will likely affect almost the entire volume of the nanoparticle. Accordingly, a single screw or edge dislocation in a Fe nanoparticle is expected to provide a significant additional contribution to the effective magnetic energy barrier, which may eventually give rise to the energy barriers required for the observed FM states. Dislocations can also explain the metastability of the FM states in Fe nanoparticles. The high mobility of dislocations as known from bulk Fe allow them to be ejected from the finite volume of nanoparticles, e.g., upon thermal excitations, as demonstrated in molecular dynamics simulations [92]. The removal of the dislocation simultaneously lowers the elastic energy stored in the lattice of the particle as well as the magneto-elastic anisotropy contribution and thus could account for a transition from FM to more bulk-like SPM behaviour.

Much less is known so far about the effect of stacking faults and twinning on the magnetic properties of Fe. Theoretical work for bcc Fe predicts an influence of the magnetism on the stacking fault energy, which suggests that there is also an effect of the defects on the magnetic properties [93]. Stacking fault energies in bcc Fe are comparably high and therefore such defects are likely metastable in nanoparticles as for dislocations. In fact, stacking faults or twinning have been observed thus far only in bulk-like systems under high mechanical stresses, but not in bcc Fe nanoparticles [16,60,67,94]. Similarly, there is currently no evidence for dislocations or other defects in bcc Fe nanoparticles, including the present HR-STEM results. Based on our experimental observations, we assign the lack of direct experimental evidence for the existence of lattice defects in Fe nanoparticles to their metastability. In particular, the FM states in Fe nanoparticles are only stabilized upon deposition onto substrates with a sufficiently low free surface energy as the present passivated Si wafers and they can spontaneously relax over time even at room temperature [40,82]. Thus, direct observation of the associated metastable defects in bcc Fe nanoparticles poses a challenging task. A successful route could be to embed the nanoparticles into suitable matrix materials to stabilize the FM states, prevent oxidation, and still allow for detailed transmission electron microscopy investigations.

For Co nanoparticles, the RHEED data also show the presence of only one crystallographic structure, the fcc lattice. This is in agreement with other reports about gas phase grown Co nanoparticles in the present size range (8 to 20 nm) [56]. For edge (screw) dislocations within a {111} slip plane in bulk Co one can estimate uniaxial anisotropy energies of 117 (68) μeV/atom at a distance of 0.5 nm from the dislocation core, which can be compared to the magnetic energy barrier of 0.3 μeV/atom given by the magneto-crystalline anisotropy. Thus, similar to Fe, these defects can significantly contribute to the total magnetic energy barrier of a nanoparticle. In contrast to bcc Fe, stacking faults and twinning are frequently observed in fcc Co nanoparticles [56,84]. Also, the present HR-STEM data show grains or twin boundaries in a number of particles, stable even upon ambient air exposure, cf. zones "A" and "B" in Fig. 3(h). Stacking faults in fcc materials can yield local hcp stacking. Based on the properties of bulk Co, hcp stacking could give rise to uniaxial anisotropies along the local c-axis with an energy barrier of 35 μeV/atom, which is also much larger when compared to the magneto-crystalline anisotropy of fcc Co. Theory shows further that a single stacking fault in Co has a long range effect on the electronic and magnetic properties of the adjacent atomic layers [95]. Thus stacking faults can also significantly contribute to the magnetic energy barriers in Co nanoparticles. Moreover, they are to first order not related with strain and the formation of local hcp stacking may even lower the cohesion energy for Co nanoparticles [56]. Stacking faults might therefore be more stable when compared to dislocations and might be even promoted by thermal annealing. If so, a growing proportion of hcp stacking in individual particles could be related with the increasing number of FM Co nanoparticles observed upon thermal annealing as shown in Fig. 5. A respective two-phase mixture of fcc and hcp stacking in individual cobalt nanoparticles was indeed reported in Ref. [96]. Their thermal behaviour may further indicate that the FM properties in the Co nanoparticles are not due to metastable dislocations, which would be ejected from the particle as discussed for bcc Fe. In fact, in metallic fcc nanoparticles combinations of different defects have been observed. For instance, in multiply twinned fcc platinum (Pt) nanoparticles a complex combination of stacking faults, screw and edge dislocations was reported [88]. In these cases it is assumed that the dislocations reduce strain which results from a geometrical mismatch of the tetrahedral building blocks and thus stabilize the total structure. The effective magnetic energy barriers in fcc Co nanoparticles would then be the result of a complex competition between different anisotropy contributions.

Also the Ni nanoparticles exhibit only fcc lattice in RHEED. Estimates of the magneto-elastic anisotropy energy of edge (screw) dislocations along the <111> direction yields 95 (55) μeV/atom at a distance of 0.5 nm from the core. This value is almost as high as for fcc Co. Since we find no FM states at RT, our data suggest that such dislocations are not stable in these particles. For fcc Ni nanoparticles, some authors have observed stacking faults or multiple twinning [32,85]. If hcp stacking is created in Ni, the present literature suggests that not only the magnetic energy barriers might be modified, but also the magnetic order could be locally affected. For hcp Ni nanoparticles, antiferromagnetic or ferromagnetic order or paramagnetic properties have been reported [26,27,30,97]. For the present Ni nanoparticles we observe FM states at 100 K. Thus these particles possess at least partial ferromagnetic order. The fact that we observe no FM states at RT suggests that stacking faults or any possible combination of defects in these particles are either not present in the investigated Ni nanoparticles or they do not yield sufficiently large magnetic energy barriers for stable room temperature magnetism.

Thus, besides magneto-crystalline, surface and shape anisotropies, lattice defects can significantly contribute to the magnetic properties of nanoparticles. The present data as well as a number of reports in the literature suggest that lattice defects in *3d* transition metal nanoparticles are abundant and thus important for the understanding of their magnetic properties. Lattice defects alter not only the magnetic anisotropy energy or exchange interaction as discussed above, but are also known to affect the local magnetic structure. For instance, in thin films it was shown that the strain fields associated with screw or edge dislocations give rise to local non-collinear spin arrangements such as vortex- or lobe-like structures, which extend up to a few nanometres around the dislocation core [98]. While such a perturbation presents only a local phenomenon in the magnetic structure of a thin film, a similar dislocation would likely modify the entire magnetic structure of a nanoparticle in the present size range. If so, defects could result in non-collinear spin structures at dimensions far below the critical sizes for which the formation of magnetic single domain states is so far expected based on the dipolar interactions [5]. This would lead to a significantly different magnetic behaviour with particular impact on the analysis of magnetization curves. In case of screw dislocations, the broken inversion symmetry of the lattice might in addition give rise to magnetic chirality effects, which could lead to novel and thus far unexplored phenomena in magnetic nanoparticles [99].

## V. CONCLUSIONS

In summary, we have studied the magnetic properties of ensembles of size- and shape-selected bcc Fe, fcc Co and fcc Ni nanoparticles with single particle sensitivity. Using a complementary microscopy approach we have directly correlated size and magnetism of a large number of individual nanoparticles. For bcc Fe and fcc Co our results clearly demonstrate that non-interacting and chemically pure nanoparticles of the same size and shape can have significantly distinct magnetic properties. Specifically, we find that a large portion of Fe and Co nanoparticles are found in a state with strikingly enhanced magnetic energy barriers manifested in magnetically blocked states at room temperature at sizes as small as 8 nm. While this unique state is irreversibly lost after thermal annealing in case of Fe, it can be promoted in the case of Co, which is promising for applications where high saturation magnetization and high magnetic anisotropy at small sizes are required. No such state is observed for fcc Ni nanoparticles at room temperature, but magnetic blocking is found at 100 K. Atomic level simulations show that effective surface and shape anisotropy contributions can lead to a sizeable enhancement of the magnetic energy barriers in all nanoparticles, but that these contributions are not sufficient to account for the observed room temperature blocking in the smallest Fe and Co nanoparticles under investigation. Similarly, their temperature-dependent behaviour cannot be explained by surface or shape effects. Based on these findings and complementary structural data we assign the strongly enhanced magnetic energy barriers and the metastable magnetic properties of the nanoparticles to lattice defects such as dislocations and stacking faults. Another likely consequence of the presence of such defects might be the occurrence of unexpected magnetic order, altered spin structures or novel properties such as magnetic chirality effects. To reveal these phenomena and to achieve an improved understanding of the magnetic properties of nanoparticles, increased experimental and theoretical efforts are urgently needed. Finally, our work underlines the importance of complementary single particle investigations for improving the understanding and control over magnetic phenomena at the nanoscale.

**ACKNOWLEDGEMENTS**

We thank A. Weber, R. Schelldorfer and J. Krbanjevic (Paul Scherrer Institut) for technical assistance. This work was supported by the Swiss Nanoscience Institute, University of Basel. A.F.R. acknowledges support from the MICIIN "Ramón y Cajal" Programme. A.B. and J.V. acknowledge funding from the European Union under the ERC Starting Grant No. 278510 and under a contract for Integrated Infrastructure Initiative ESTEEM2 No. 312483. R.Y. and U.N. thank the Deutsche Forschungsgemeinschaft for financial support via SFB 1214. Part of this work was performed at the Surface/ Interface:Microscopy (SIM) beam line of the Swiss Light Source, Paul Scherrer Institut, Villigen, Switzerland.

# REFERENCES

[1]     A. H. Lu, E. L. Salabas, and F. Schuth, Angew. Chem. Int. Ed. **46**, 1222 (2007).

[2]     J. Bansmann *et al.*, Surf. Sci. Rep. **56**, 189 (2005).

[3]     N. Jones, Nature (London) **472**, 22 (2011).

[4]     B. Q. Geng, Z. L. Ding, and Y. Q. Ma, Nano Res. **9**, 2772 (2016).

[5]     A. Aharoni, *Introduction to the Theory of Ferromagnetism* (Clarendon, Oxford, 1996).

[6]     S. Sun, C. B. Murray, D. Weller, L. Folks, and A. Moser, Science **287**, 1989 (2000).

[7]     V. Skumryev, S. Stoyanov, Y. Zhang, G. Hadjipanayis, D. Givord, and J. Nogues, Nature (London) **423**, 850 (2003).

[8]     I. M. L. Billas, A. Chatelain, and W. A. de Heer, Science **265**, 1682 (1994).

[9]     R. H. Kodama, J. Magn. Magn. Mater. **200**, 359 (1999).

[10]    C. Antoniak *et al.*, Nat. Commun. **2**, 528 (2011).

[11]    P. Andreazza, V. Pierron-Bohnes, F. Tournus, C. Andreazza-Vignolle, and V. Dupuis, Surf. Sci. Rep. **70**, 188 (2015).

[12]    M. Estrader *et al.*, Nanoscale **7**, 3002 (2015).

[13]    J. Carvell, E. Ayieta, A. Gavrin, R. H. Cheng, V. R. Shah, and P. Sokol, J. Appl. Phys. **107**, 103913 (2010).

[14]    F. Bodker, S. Morup, and S. Linderoth, Phys. Rev. Lett. **72**, 282 (1994).

[15]    F. Kronast *et al.*, Nano Lett. **11**, 1710 (2011).

[16]    M. Jamet, W. Wernsdorfer, C. Thirion, V. Dupuis, P. Melinon, A. Perez, and D. Mailly, Phys. Rev. B **69**, 024401 (2004).

[17]    D. L. Peng, T. Hihara, K. Sumiyama, and H. Morikawa, J. Appl. Phys. **92**, 3075 (2002).

[18]    J. P. Pierce, M. A. Torija, Z. Gai, J. Shi, T. C. Schulthess, G. A. Farnan, J. F. Wendelken, E. W. Plummer, and J. Shen, Phys. Rev. Lett. **92**, 237201 (2004).

[19]   A. V. Trunova, R. Meckenstock, I. Barsukov, C. Hassel, O. Margeat, M. Spasova, J. Lindner, and M. Farle, J. Appl. Phys. **104**, 093904 (2008).

[20]   J. P. Chen, C. M. Sorensen, K. J. Klabunde, and G. C. Hadjipanayis, J. Appl. Phys. **76**, 6316 (1994).

[21]   M. Respaud, J. M. Broto, H. Rakoto, A. R. Fert, L. Thomas, B. Barbara, M. Verelst, E. Snoeck, P.Lecante, A. Mosset, J. Osuna, T. O. Ely, C. Amiens, and B. Chaudret, Phys. Rev. B **57**, 2925 (1998).

[22]   F. Luis, J. M. Torres, L. M. García, J. Bartolomé, J. Stankiewicz, F. Petroff, F. Fettar, J. L. Maurice, and A. Vaurès, Phys. Rev. B **65**, 094409 (2002).

[23]   S. Oyarzun, A. Tamion, F. Tournus, V. Dupuis, and M. Hillenkamp, Sci. Rep. **5**, 14749 (2015).

[24]   R. Morel, A. Brenac, C. Portemont, T. Deutsch, and L. Notin, J. Magn. Magn. Mater. **308**, 296 (2007).

[25]   N. Weiss, T. Cren, M. Epple, S. Rusponi, G. Baudot, S. Rohart, A. Tejeda, V. Repain, S. Rousset, P. Ohresser, F. Scheurer, P. Bencok, and H. Brune, Phys. Rev. Lett. **95**, 157204 (2005).

[26]   M. Han, Q. Liu, J. H. He, Y. Song, Z. Xu, and J. M. Zhu, Adv. Mater. **19**, 1096 (2007).

[27]   Y. T. Jeon, J. Y. Moon, G. H. Lee, J. Park, and Y. M. Chang, J. Phys. Chem. B **110**, 1187 (2006).

[28]   T. Ould-Ely, C. Amiens, B. Chaudret, E. Snoeck, M. Verelst, M. Respaud, and J. M. Broto, Chem. Mat. **11**, 526 (1999).

[29]   N. Cordente, M. Respaud, F. Senocq, M. J. Casanove, C. Amiens, and B. Chaudret, Nano Lett. **1**, 565 (2001).

[30]   D. X. Chen, O. Pascu, and A. Roig, J. Magn. Magn. Mater. **363**, 195 (2014).

[31]     M. Maicas, M. Sanz, H. Cui, C. Aroca, and P. Sanchez, J. Magn. Magn. Mater. **322**, 3485 (2010).

[32]     E. Kita, N. Tsukuhara, H. Sato, K. Ota, H. Yangaihara, H. Tanimoto, and N. Ikeda, Appl. Phys. Lett. **88**, 152501 (2006).

[33]     S. Rohart, V. Repain, A. Tejeda, P. Ohresser, F. Scheurer, P. Bencok, J. Ferré, and S. Rousset, Phys. Rev. B **73**, 165412 (2006).

[34]     M. Ruano *et al.*, Phys. Chem. Chem. Phys. **15**, 316 (2013).

[35]     S. A. Majetich and M. Sachan, J. Phys. D: Appl. Phys. **39**, R407 (2006).

[36]     J. Rockenberger, F. Nolting, J. Luning, J. Hu, and A. P. Alivisatos, J. Chem. Phys. **116**, 6322 (2002).

[37]     A. Fraile Rodríguez, F. Nolting, J. Bansmann, A. Kleibert, and L. J. Heyderman, J. Magn. Magn. Mater. **316**, 426 (2007).

[38]     A. Fraile Rodríguez, A. Kleibert, J. Bansmann, A. Voitkans, L. J. Heyderman, and F. Nolting, Phys. Rev. Lett. **104**, 127201 (2010).

[39]     C. A. F. Vaz, A. Balan, F. Nolting, and A. Kleibert, Phys. Chem. Chem. Phys. **16**, 26624 (2014).

[40]     A. Balan, P. M. Derlet, A. F. Rodríguez, J. Bansmann, R. Yanes, U. Nowak, A. Kleibert, and F. Nolting, Phys. Rev. Lett. **112**, 107201 (2014).

[41]     A. Fraile Rodríguez, A. Kleibert, J. Bansmann, and F. Nolting, J. Phys. D: Appl. Phys. **43**, 474006 (2010).

[42]     See Supplemental Material at [URL will be inserted by publisher] for additional information on the sample design, the experimental geometry in X-PEEM, the experimentally observed and simulated magnetic contrast distribution, and the model calcualtions.

[43]     R. P. Methling, V. Senz, E. D. Klinkenberg, T. Diederich, J. Tiggesbaumker, G. Holzhuter, J. Bansmann, and K. H. Meiwes-Broer, Eur. Phys. J. D **16**, 173 (2001).

[44]    A. Kleibert, J. Passig, K. H. Meiwes-Broer, M. Getzlaff, and J. Bansmann, J. Appl. Phys. **101**, 114318 (2007).

[45]    J. Passig, K.-H. Meiwes-Broer, and J. Tiggesbäumker, Rev. Sci. Instrum. **77**, 093304 (2006).

[46]    H. Haberland, Z. Insepov, and M. Moseler, Phys. Rev. B **51**, 11061 (1995).

[47]    V. N. Popok, I. Barke, E. E. B. Campbell, and K.-H. Meiwes-Broer, Surf. Sci. Rep. **66**, 347 (2011).

[48]    A. Kleibert, A. Voitkans, and K. H. Meiwes-Broer, Phys. Rev. B **81**, 073412, 073412 (2010).

[49]    A. Kleibert, A. Voitkans, and K. H. Meiwes-Broer, Phys. Status Solidi B **247**, 1048 (2010).

[50]    S. Bartling *et al.*, ACS Nano **9**, 5984 (2015).

[51]    L. Le Guyader, A. Kleibert, A. Fraile Rodríguez, S. El Moussaoui, A. Balan, M. Buzzi, J. Raabe, and F. Nolting, J. Electron. Spectrosc. Relat. Phenom. **185**, 371 (2012).

[52]    J. Stohr, Y. Wu, B. D. Hermsmeier, M. G. Samant, G. R. Harp, S. Koranda, D. Dunham, and B. P. Tonner, Science **259**, 658 (1993).

[53]    H. Hövel and I. Barke, Prog. Surf. Sci. **81**, 53 (2006).

[54]    F. Baletto and R. Ferrando, Rev. Mod. Phys. **77**, 371 (2005).

[55]    V. F. Puntes, K. M. Krishnan, and A. P. Alivisatos, Science **291**, 2115 (2001).

[56]    O. Kitakami, H. Sato, Y. Shimada, F. Sato, and M. Tanaka, Phys. Rev. B **56**, 13849 (1997).

[57]    S. H. Baker, M. Roy, S. J. Gurman, S. Louch, A. Bleloch, and C. Binns, J. Phys.: Condens. Matter **16**, 7813 (2004).

[58]    T. Ling, J. Zhu, H. M. Yu, and L. Xie, J. Phys. Chem. C **113**, 9450 (2009).

[59]    G. Wulff, Z. Krystallogr. **34**, 449 (1901).

[60]   T. Vystavel, G. Palasantzas, S. A. Koch, and J. T. M. De Hosson, Appl. Phys. Lett. **82**, 197 (2003).

[61]   C. Mottet, J. Goniakowski, F. Baletto, R. Ferrando, and G. Treglia, Phase Transit. **77**, 101 (2004).

[62]   A. Kleibert, W. Rosellen, M. Getzlaff, and J. Bansmann, Beilstein J. Nanotechnol. **2**, 47 (2011).

[63]   Y. Kobayashi, Y. Shinoda, and K. Sugii, Jpn. J. Appl. Phys. **29**, 1004 (1990).

[64]   F. Masini, P. Hernandez-Fernandez, D. Deiana, C. E. Strebel, D. N. McCarthy, A. Bodin, P. Malacrida, I. Stephens, and I. Chorkendorff, Phys. Chem. Chem. Phys. **16**, 26506 (2014).

[65]   A. Kleibert, F. Bulut, R. K. Gebhardt, W. Rosellen, D. Sudfeld, J. Passig, J. Bansmann, K. H. Meiwes-Broer, and M. Getzlaff, J. Phys.: Condens. Matter **20**, 445005 (2008).

[66]   J. Nogues, V. Skumryev, J. Sort, S. Stoyanov, and D. Givord, Phys. Rev. Lett. **97**, 157203 (2006).

[67]   A. Pratt, L. Lari, O. Hovorka, A. Shah, C. Woffinden, S. P. Tear, C. Binns, and R. Kroger, Nat. Mater. **13**, 26 (2014).

[68]   D. A. Garanin and H. Kachkachi, Phys. Rev. Lett. **90**, 065504 (2003).

[69]   R. Yanes, O. Chubykalo-Fesenko, H. Kachkachi, D. A. Garanin, R. Evans, and R. W. Chantrell, Phys. Rev. B **76**, 064416, 064416 (2007).

[70]   H. Kachkachi and E. Bonet, Phys. Rev. B **73**, 224402 (2006).

[71]   F. Garcia-Sanchez and O. Chubykalo-Fesenko, Appl. Phys. Lett. **93** (2008).

[72]   L. S. Campana, A. Caramico D'Auria, M. D'Ambrosio, U. Esposito, L. De Cesare, and G. Kamieniarz, Phys. Rev. B **30**, 2769 (1984).

[73]   U. Nowak, O. N. Mryasov, R. Wieser, K. Guslienko, and R. W. Chantrell, Phys. Rev. B **72**, 172410 (2005).

[74]    J. G. Gay and R. Richter, Phys. Rev. Lett. **56**, 2728 (1986).

[75]    X. Qian and W. Hubner, Phys. Rev. B **64**, 092402 (2001).

[76]    J. Ye, W. He, Q. Wu, H. L. Liu, X. Q. Zhang, Z. Y. Chen, and Z. H. Cheng, Sci. Rep. **3**, 2148 (2013).

[77]    H. L. Liu, W. He, Q. Wu, J. Ye, X. Q. Zhang, H. T. Yang, and Z. H. Cheng, AIP Adv. **3**, 062101 (2013).

[78]    A. Kharmouche, S. M. Cherif, A. Bourzami, A. Layadi, and G. Schmerber, J. Phys. D: Appl. Phys. **37**, 2583 (2004).

[79]    P. Poulopoulos and K. Baberschke, J. Phys.: Condens. Matter **11**, 9495 (1999).

[80]    M. Beleggia, M. De Graef, and Y. Millev, Philos. Mag. **86**, 2451 (2006).

[81]    C. A. F. Vaz, J. A. C. Bland, and G. Lauhoff, Rep. Prog. Phys. **71**, 056501 (2008).

[82]    A. Balan, A. F. Rodriguez, C. A. F. Vaz, A. Kleibert, and F. Nolting, Ultramicroscopy **159**, 513 (2015).

[83]    I. Barke *et al.*, Nat. Commun. **6** (2015).

[84]    W. Wernsdorfer, C. Thirion, N. Demoncy, H. Pascard, and D. Mailly, J. Magn. Magn. Mater. **242**, 132 (2002).

[85]    S. D'Addato, V. Grillo, S. Altieri, R. Tondi, S. Valeri, and S. Frabboni, J. Phys.: Condens. Matter **23**, 175003 (2011).

[86]    S. H. Baker, M. Roy, S. C. Thornton, and C. Binns, J. Phys.: Condens. Matter **24**, 176001 (2012).

[87]    A. Mayoral, H. Barron, R. Estrada-Salas, A. Vazquez-Duran, and M. Jose-Yacaman, Nanoscale **2**, 335 (2010).

[88]    C. C. Chen, C. Zhu, E. R. White, C. Y. Chiu, M. C. Scott, B. C. Regan, L. D. Marks, Y. Huang, and J. W. Miao, Nature (London) **496**, 74 (2013).

[89]    L. D. Marks, Rep. Prog. Phys. **57**, 603 (1994).

[90]     H. Kronmüller and M. Fähnle, *Micromagnetism and the microstructure of ferromagnetic solids* (Cambridge University Press, Cambridge, 2003), Cambridge studies in magnetism.

[91]     D. Pomfret and M. Prutton, Phys. Status Solidi A **19**, 423 (1973).

[92]     H. B. Liu, G. Canizal, S. Jimenez, M. A. Espinosa-Medina, and J. A. Ascencio, Comp. Mater. Sci. **27**, 333 (2003).

[93]     J. A. Yan, C. Y. Wang, and S. Y. Wang, Phys. Rev. B **70**, 174105 (2004).

[94]     T. L. Altshuler and J. W. Christian, Acta Metall. **14**, 903 (1966).

[95]     C. J. Aas, L. Szunyogh, R. F. L. Evans, and R. W. Chantrell, J. Phys.: Condens. Matter **25** (2013).

[96]     V. Dureuil, C. Ricolleau, M. Gandais, and C. Grigis, Eur. Phys. J. D **14**, 83 (2001).

[97]     R. Masuda *et al.*, Sci. Rep. **6**, 20861 (2016).

[98]     L. Berbil-Bautista, S. Krause, M. Bode, A. Badía-Majós, C. de la Fuente, R. Wiesendanger, and J. I. Arnaudas, Phys. Rev. B **80**, 241408 (2009).

[99]     A. B. Butenko and U. K. Rossler, EPJ Web. Conf. **40**, 08006 (2013).

SUPPLEMENTARY MATERIAL

**Direct observation of enhanced magnetism in individual size- and shape-selected *3d* transition metal nanoparticles**

Armin Kleibert[*,1], Ana Balan[1], Rocio Yanes[2], Peter M. Derlet[3], Carlos A. F. Vaz[1], Martin Timm[1], Arantxa Fraile Rodríguez[4], Armand Béché[5], Jo Verbeeck[5], Rajen S. Dhaka[1,6,7], Milan Radovic[1,7,8], Ulrich Nowak[2], and Frithjof Nolting[1]

[1]Swiss Light Source, Paul Scherrer Institut, 5232 Villigen PSI, Switzerland

[2]Department of Physics, University of Konstanz, 78457 Konstanz, Germany

[3]Condensed Matter Theory Group, NUM, Paul Scherrer Institut, 5232 Villigen PSI, Switzerland

[4]Departament de Física de la Matèria Condensada and Institut de Nanociència i Nanotecnologia (IN2UB), Universitat de Barcelona, 08028 Barcelona, Spain

[5]EMAT, University of Antwerp, 2020 Antwerpen, Belgium

[6]Department of Physics, Indian Institute of Technology Delhi, Hauz Khas, New Delhi-110016, India

[7]Institute of Condensed Matter Physics, Ecole Polytechnique Fédérale de Lausanne (EPFL), 1015 Lausanne, Switzerland

[8]SwissFEL, Paul Scherrer Institut, 5232 Villigen PSI, Switzerland

*Corresponding author, e-mail: armin.kleibert@psi.ch

## 1. Substrates with Au-Markers for Complementary Microscopy

To enable identification of the very same nanoparticles using complementary microscopy, gold markers are prepared on Si(001) wafers by means of electron beam lithography. The appearance of the markers (here: "C1") and the nanoparticles in the different microscopes are shown in Fig. S-1.

## 2. X-PEEM Experimental Geometry

In X-PEEM the samples are illuminated with polarized, monochromatic synchrotron radiation with a propagation vector $\vec{k}$ impinging at an angle of incidence $\theta_k = 16°$ with respect to the sample surface as shown in Fig. S-2. A raw X-PEEM image is shown in Fig. S-1(a). Using circularly polarized synchrotron radiation the X-ray magnetic circular dichroism (XMCD) effect leads to a magnetization-dependent intensity of a nanoparticle with a magnetic moment $\vec{m} = \vec{m}(\theta_m, \phi_m)$ given by $I(C^{\pm}) = I_0 \pm \gamma \vec{k} \cdot \vec{m}$, where $I_0$ is the isotropic (i. e. non-magnetic) intensity, $\gamma$ is a material and photon energy dependent constant, and $C^{\pm}$ denotes circular right- and left-handed polarization.

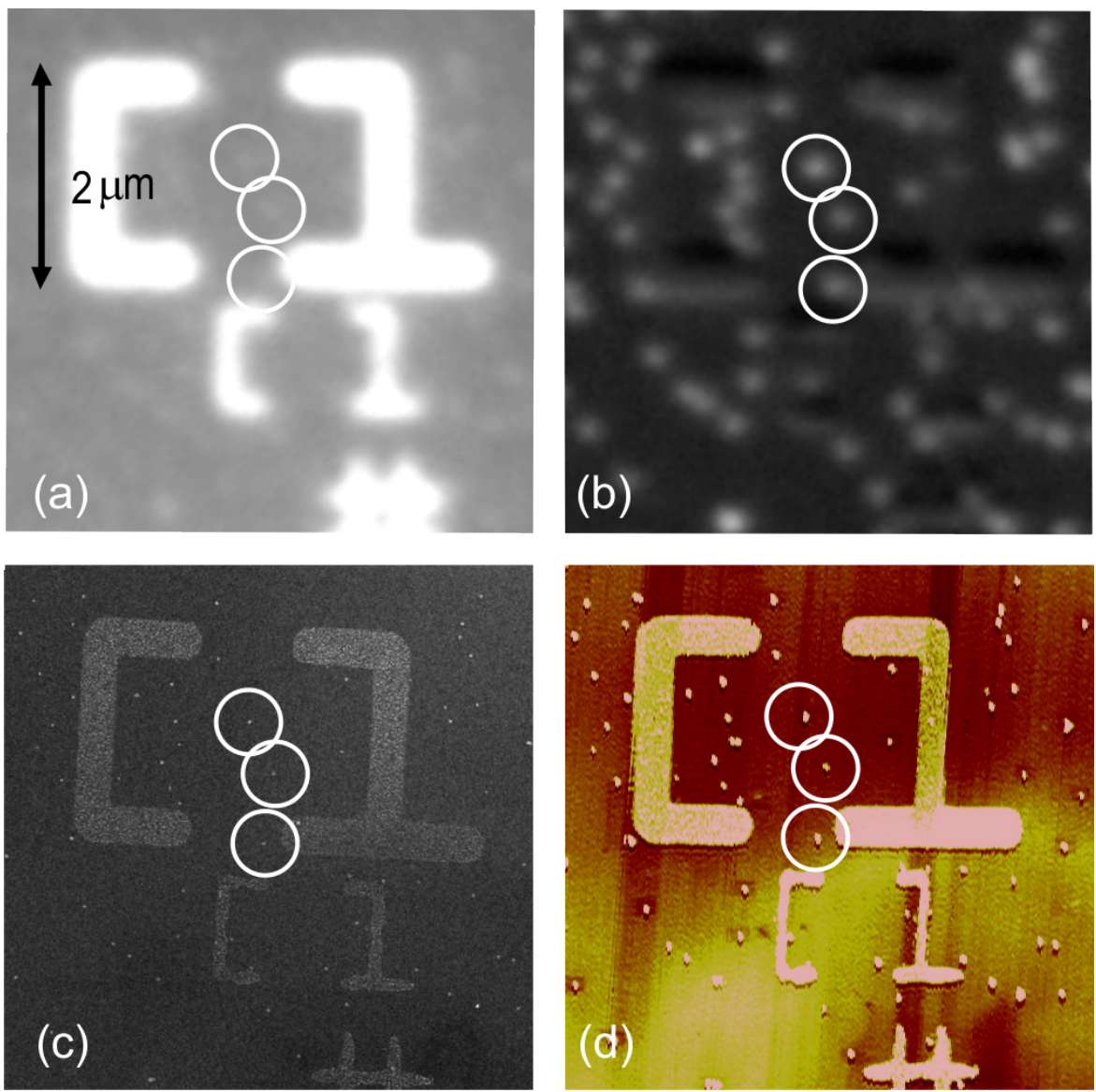

**Figure S-1.** Si wafer substrates with Au-marker structure and Fe nanoparticles in (a) raw X-PEEM image, (b) elemental contrast image obtained at the Fe $L_3$ edge, (c) SEM image, and (d) AFM image. To illustrate the particle identification, three nanoparticles are highlighted with circles.

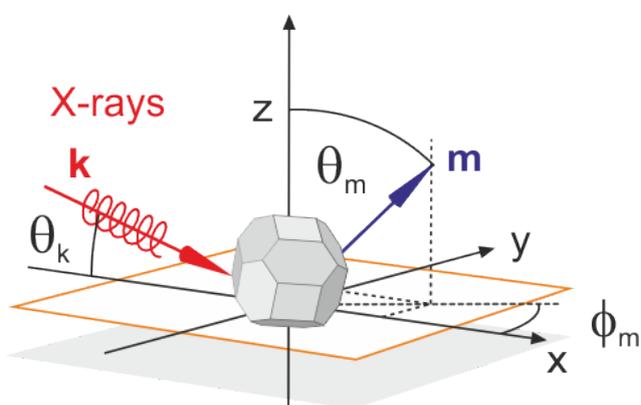

**Figure S-2.** Experimental geometry in X-PEEM.

## 3. Experimentally Observed Magnetic Contrast Distribution

A quantitative analysis of the magnetic contrast distribution is obtained by calculating the XMCD asymmetry $A = (C^+ - C^-)/(C^+ + C^-) = \gamma \vec{k} \cdot \vec{m}/I_0$ of each of the selected individual nanoparticles. Experimentally obtained histograms of the respective XMCD asymmetry for the different systems for the Fe, Co, and Ni nanoparticle samples in their *as grown* state are shown in Fig. S-3. A negative (positive) value of the asymmetry $A$ corresponds to a dark (bright) grey contrast in the respective magnetic contrast maps, see Figs. 1(d) and 1(e). As discussed in Ref. [1], the central peak with $|A| \leq 0.03$ corresponds to nanoparticles without magnetic contrast and can be assigned to SPM nanoparticles, while the flat part of the distribution is due to magnetic blocked FM nanoparticles. Figs. S-3(a) and S-3(b) show that Fe and Co nanoparticles have almost the same contributions of SPM and FM fractions and similar form of histograms in their as-deposited state, while the Ni nanoparticles in Fig. S-3(c) exhibit only the SPM peak.

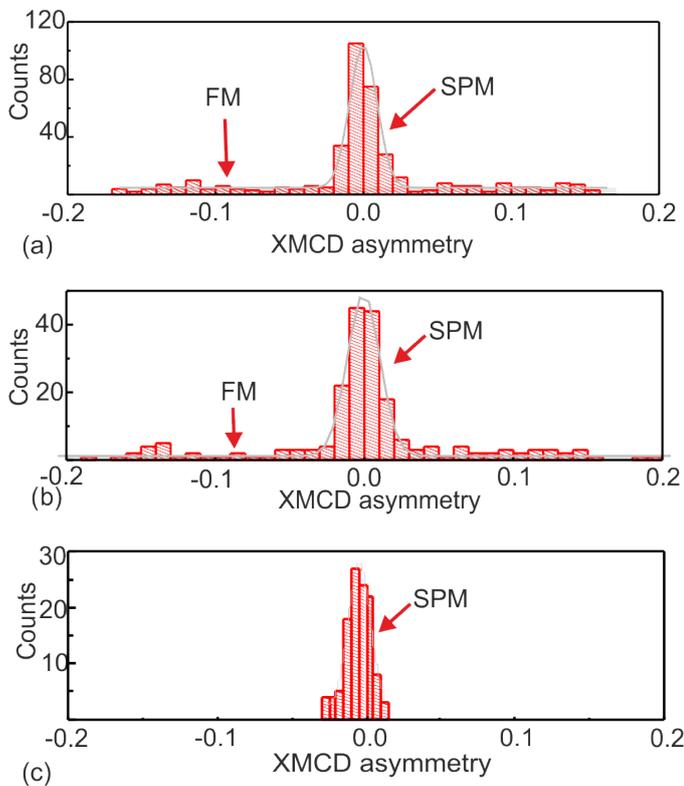

**Figure S-3.** Experimentally obtained histograms of XMCD asymmetry of (a) Fe, (b) Co, and (c) Ni nanoparticles in the *as grown* state.

## 3. Simulated Magnetic Contrast Distribution for Different Scenarios

In order to assess the experimentally observed flat distribution of the normalized asymmetries of the FM Fe and Co nanoparticles, we have simulated XMCD asymmetry distributions resulting from three different scenarios for the orientation distribution of the magnetic moments of the deposited nanoparticles. In these simulations, the expected XMCD asymmetry is calculated for each nanoparticle according to the experimental geometry and the orientation of its magnetic moment given by a unit vector in spherical coordinates $(\theta_m, \phi_m)$, cf. Fig. S-2. The simulations show that an ensemble of nanoparticles with fully random orientation of their magnetic moments results in a flat distribution of the XMCD asymmetry, cf. Fig. S-4(a). An in-plane random orientation of the magnetic moments results in the distribution shown in Fig. S-4(b). Finally, Fig. S-4(c) shows the histogram for a nanoparticle ensemble with preferred outofplane magnetization, where we allowed a standard deviation of 45° from full out of plane orientation ($\theta_m = 0°$ or $\theta_m = 180°$, respectively) of the moment to mimic a sample with deviations from perfect out-of-plane alignment of the magnetic moments. Figs. S-4(a) - S-4(c) show that the three scenarios result in a distinct distribution of the XMCD asymmetry. It follows that only a fully random orientation of magnetic moments yields a flat histogram compatible with the experimental data for the FM portion of the Fe and Co nanoparticles as shown in Figs. S-3(a) and S-3(b).

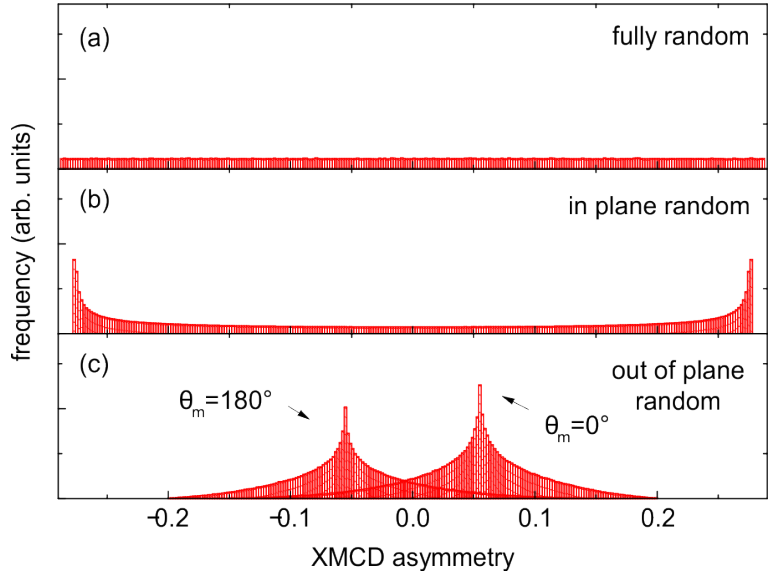

**Figure S-4.** Simulated XMCD asymmetry distribution for (a) fully, (b) in-plane, and (c) outofplane randomly orientated magnetic moments according to the experimental geometry shown in Fig. S-2. Each simulation was performed for $10^6$ particles.

## 4. Parameters Used for the Calculations of the Magnetic Energy Barriers

| Material | bcc Fe | fcc Co | fcc Ni |
|---|---|---|---|
| Lattice parameter $a$ [Å] | 2.87 | 3.548 | 3.52 |
| Magnetic moment $\mu$ [$\mu_B$/atom] | 2.17 | 1.72 | 0.6 |
| Saturation magnetization $M_s$ [emu/cm$^3$] | 1702.59 | 1428.58 | 510.329 |
| First order anisotropy constant $K_1$ [erg/cm$^3$] | 4.8×10$^5$ | -7.14×10$^5$ | -5.7×10$^4$ |
| Second order anisotropy constant $K_2$ [erg/cm$^3$][a] | 0 | 0 | -2.3×10$^4$ |
| Exchange constant $J$ [10$^{-14}$ erg] | 6.8769 | 6.000 | 2.6988 |
| Bulk Curie temperature $T_c$ [K] | 1044 | 1403 | 624-631 (exp.) |

[a] We have set $K_2 = 0$ erg/cm$^3$ in the calculations for bcc Fe and fcc Co, since they are small compared to the respective values of $K_1$.

# 5. Magnetic Energy Barriers Required for a Magnetically Blocked State at Room Temperature

The temperature-depending switching rate $\tau_S$ of a magnetic nanoparticle can be expressed by an Arrhenius law:

$$\tau_s = \frac{1}{v_0} e^{E_m / k_B T}$$

where $v_0$ is the attempt frequency, $E_m$ is the magnetic energy barrier and $k_B$ is the Boltzmann constant. A magnetically blocked state is observed when the switching rate $\tau_s$ becomes equal or smaller than the measurement time $\tau_x$. To estimate $E_m$, the attempt frequency has to be determined.

The attempt frequency in the case of cubic anisotropy with $K_1 > 0$ can be obtained by a relation used by Coffey and Kalmykov in Ref. [2] in the limit of high damping ($\alpha \geq 1$):

$$v_0 = \frac{4\sqrt{2} E_m (\alpha + \sqrt{9\alpha + 8}) \gamma}{M_s V (1 + \alpha^2) \pi}$$

For cubic anisotropy with $K_1 < 0$:

$$v_0 = \frac{4\sqrt{2} E_m (\sqrt{9\alpha + 8} - \alpha) \gamma}{M_s V (1 + \alpha^2) \pi}$$

Here, $\gamma$ is the gyromagnetic ratio of an isolated electron being $1.760 \times 10^{11}$ rad×s$^{-1}$×T$^{-1}$ and $V$ is the volume of the particle. To evaluate these equations, we consider further the temperature-dependence of the saturation magnetization:

$$M_s(T) = M_0 \sqrt{1 - \frac{T}{T_c}}$$

Where $T_c$ is the Curie temperature and $M_0$ the saturation magnetization at $T = 0$ K. A lower limit for the attempt frequency is obtained using the parameters for bulk bcc Fe, fcc Co and fcc Ni considering spherical nanoparticles with a diameter of $D = 20$ nm. The damping coefficient is set to $\alpha = 1$. At room temperature we obtain $v_0 = 6.3 \times 10^9$ s$^{-1}$ for bcc Fe, $v_0$

=$1.9\times10^9$ s$^{-1}$ for fcc Co, and $v_0$ = $4.6\times10^8$ s$^{-1}$ for fcc Ni nanoparticles. These values yield the following minimum magnetic energy barriers for a magnetically blocked state with $\tau_S = \tau_x$: $E_m = 0.68$ eV for bcc Fe, $E_m = 0.67$ eV for fcc Co, $E_m = 0.71$ eV for fcc Ni as shown by the dashed lines in Fig. 6 of the manuscript.

[1] Balan, A.; Derlet, P. M.; Fraile Rodríguez, A.; Bansmann, J.; Yanes, R.; Nowak, U.; Kleibert, A.; Nolting, F. Direct observation of magnetic metastability in individual iron nanoparticles. *Phys. Rev. Lett.* **2014**, *112*, 107201.

[2] Coffey, W. T.; Kalmykov, Y. P. Thermal fluctuations of magnetic nanoparticles: Fifty years after Brown. *J. Appl. Phys.* **2012**, *112*, 121301.